# Optically Driven Gold Nanoparticles Seed Surface Bubble Nucleation in Plasmonic Suspension

Qiushi Zhang[1], Ruiyang Li[1], Eungkyu Lee[1,4*], and Tengfei Luo[1,2,3*]

[1]Department of Aerospace and Mechanical Engineering, University of Notre Dame, Notre Dame, IN, USA

[2]Department of Chemical and Biomolecular Engineering, University of Notre Dame, Notre Dame, USA

[3]Center for Sustainable Energy of Notre Dame (ND Energy), University of Notre Dame, Notre Dame, USA

[4]Department of Mechanical System Engineering, Kumoh National Institute of Technology, 61 Daehak-ro, Gumi, Gyeongbuk, 730-701, South Korea

*Correspondence to: tluo@nd.edu; elee@kumoh.ac.kr

**ABSTRACT**

Photothermal surface bubbles play important roles in a wide range of applications like catalysis, microfluidics and biosensing, but their formation on a transparent substrate immersed in a plasmonic nanoparticle (NP) suspension has an unknown origin. Here, we show that NPs deposited on the substrate by dispersive optical forces are responsible for the nucleation of such photothermal surface bubbles. High-speed videography shows that the surface bubble formation is always preceded by the optically driven NPs moving toward and adhering to the surface. We observe that the thresholds of laser power density to form a surface bubble drastically differ depending on if the surface is forward- or backward-facing the light propagation direction, which cannot be explained by a purely thermal process (e.g., volumetric heating by plasmonic NPs). This can be attributed to different optical power densities needed to enable optical pulling and pushing of NPs in the suspension. Optical pulling requires higher light intensity to excite supercavitation around NPs to enable proper optical configuration. Eventually, these optically deposited NPs work as a surface photothermal heater, seeding the surface bubble nucleation. We also find that there is a critical number density of deposited NPs to nucleate a surface bubble at a given laser power density, and it is nearly the same for both the forward- and the backward-facing surfaces. Our finding reveals interesting physics leading to photothermal surface bubble generation in plasmonic NP suspensions.

## INTRODUCTION

Surface bubbles generated by photothermal effects are playing significant roles in a wide range of applications, such as micro-bubble logics(*1*), vapor generation(*2–4*), cancer therapy(*5–9*), plasmon-assisted catalysis(*10–13*) and nanoparticle (NP) manipulation(*14–17*) and deposition(*18–20*). To realize the photothermal effect, a light-absorbing substrate, such as silicon, thin-metallic layer and conducting oxide, is usually immersed in liquid to convert optical energy into thermal energy(*15, 16, 18*). When the light intensity is sufficiently high to raise the temperature of the substrate above a threshold, a surface bubble can nucleate.

For photothermal conversion, metallic nanostructures are among the most efficient transducers, as they can support the surface plasmonic resonance (SPR) to amplify the light intensity at the metal/dielectric interface by orders of magnitude(*21, 22*). In addition, since the resonant wavelength of the SPR can be tuned by properly designing the shape, spacing and size of the metallic nanostructures at a sub-wavelength scale, there have been systematic studies of surface bubble formation with SPR (i.e., plasmonic surface bubble)(*11, 23–29*). Fundamental studies have focused on the growth dynamics of the plasmonic surface bubbles, revealing interesting physics about bubble oscillation, vaporization, and gas expelling(*30, 31*). With well-defined surface photothermal heat flux, bubble nucleation time is found to be inversely proportional to the concentration of dissolved air in liquid(*30, 32, 33*). On the periodic-metal nanostructures (i.e., pillar or cylinder), it is observed that the laser power density threshold for bubble nucleation depends on the number density of the nanostructures(*27, 34*). Overall, it has been known that the photothermal bubble generation process at a surface with deposited SPR-supporting nanostructures is in principle similar to the conventional pool boiling, where the pre-decorated substrate is working as a heat source as well as providing nucleation sites for bubble nucleation(*11, 27, 30–33, 35–40*).

However, it has been reported that surface bubble can be generated on optically transparent substrates when it is immersed in plasmonic NP suspensions(*19, 20, 29, 41*). While the plasmonic NPs suspending in liquid can support the SPR when illuminated by a resonant light to heat up the irradiated volume of liquid, that means there are no light-absorbing materials on the surface that convert optical energy to surface heating. The fundamental question here is how surface bubbles can be formed with the absence of direct surface heating source? In this work, we investigate the origin of surface bubbles generation in plasmonic NP suspensions on the transparent substrate. Interestingly, our experiments find that the thresholds of laser power density to form a surface bubble drastically differ depending on if the surface is forward-facing (FF) or backward-facing (BF) the light propagation direction. High-speed videography reveals that the light-guided NP deposition on the surface is a necessity for bubble nucleation, and it is the optical dispersive force that drives such deposition. Optical pulling force is needed to deposit NPs on the BF surface, and this is achieved only when the laser intensity is sufficiently high to generate a supercavitating nanobubble around the NP, which is necessary to enable the proper optical condition for optical pulling motion to happen(*42*). On the FF surface, the optical pushing force, which exists on both bare or supercavitating NPs, makes the NP deposition easier and thus lowers the surface bubble nucleation threshold. Further experiments comparing NPs with different SPR frequencies (e.g., core-shell (CS) and solid Au NPs) show that optical pulling-induced nucleation is only possible if the laser is at the SPR peak of the NPs, since it can intensely excite the NP to form supercavitation. On the contrary, optical pushing-induced surface bubble nucleation always happens regardless if the NP SPR peak aligns with the incident laser wavelength. These results reveal interesting physics leading to photothermal surface bubble generation in NP suspensions.

**RESULTS AND DISCUSSION**

We first demonstrate that surface bubbles can be generated in the CS NP suspension (concentration ~ $2\times10^{15}$ particles/$m^3$) when the laser at the SPR peak frequency is focused on either the BF or the FF surface with the optical system(*28, 29*) shown in **Figure 1a** (see Methods section for details, Supporting Movies M1 and M2). To shed light on the mechanism of bubble formation, we investigated the nucleation time as a function of the laser power density. The laser power density we refer to in this work is the maximum laser power density at the center of the Gaussian beam for a given laser power. We note that the nucleation time in this study is defined as the time between the moment of turning on the laser and the onset of nucleation. In the experiments, the onset of surface bubble nucleation can be identified by observing the strongly scattered light at the surface where the laser beam is focused on (see Supporting Information SI1). We note that the surface bubble nucleation in NP suspension includes a NP deposition stage after turning on the laser (as discussed later). As a result, the time interval between turning on the laser and the observation of the strong scattered light is the total time including NP deposition and the nucleation of a surface bubble. We also note that the definition of bubble nucleation time can vary in different context. For example, on prefabricated plasmonic substrates,(*30, 33*) the nucleation time is the interval between switching on the laser and the onset of a giant vapor bubble. In these cases, the giant bubble collapses within ~200 μs, followed by a subsequent oscillating bubble lasting for <2000 μs, which precedes the emergence of a stably growing vapor bubble. Since our measured nucleation times are on the order of 1~100 seconds, whether the onset of nucleation is chosen to be the start of the initial giant bubble or the subsequent stable bubble is not important in our analyses.

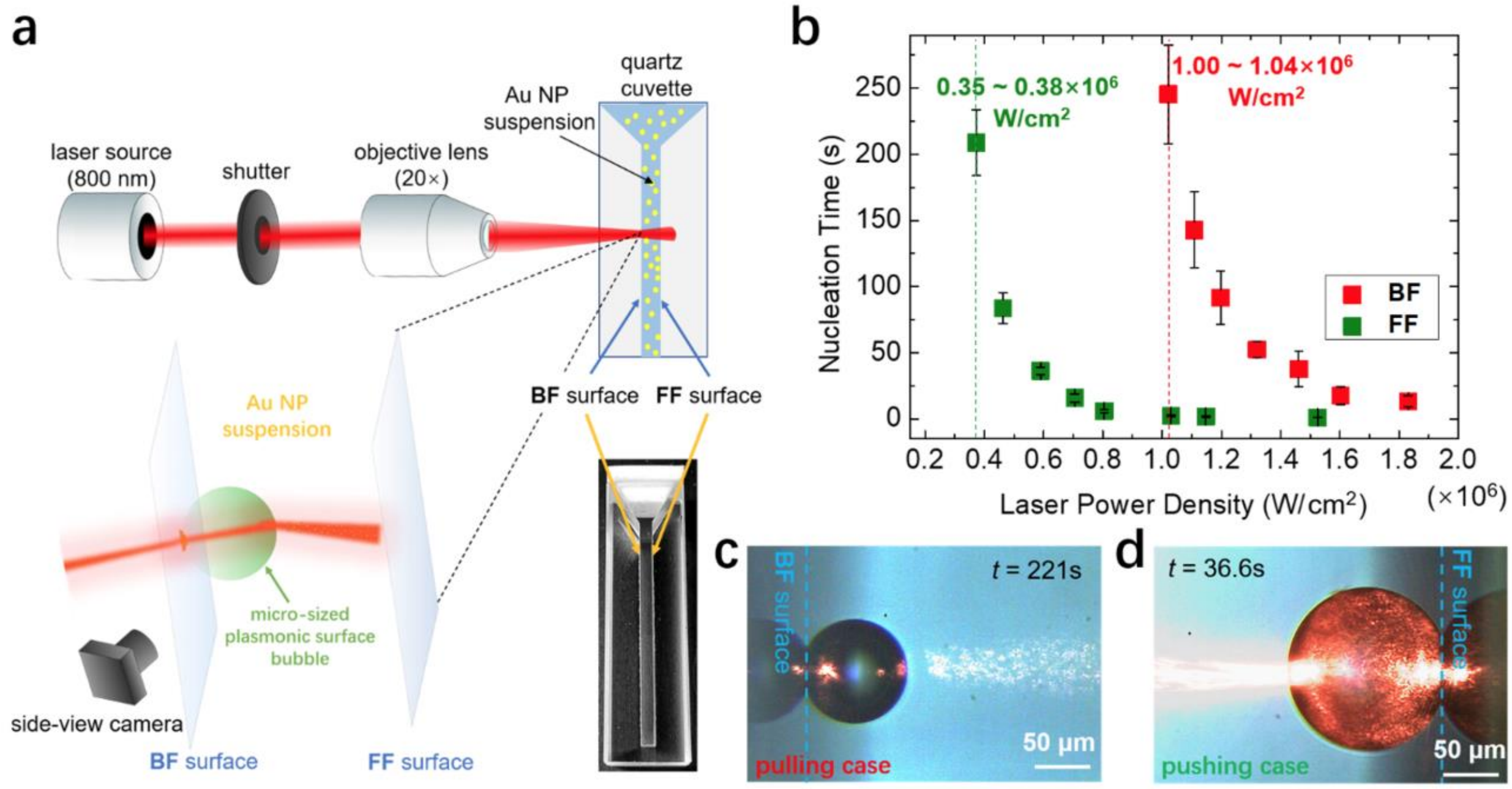

**Figure 1.** (a) Schematic of the experimental setup to characterize the surface bubble nucleation on the backward-facing (BF) or the forward-facing (FF) surface with respect to the laser propagating direction. (b) The nucleation time of surface bubble on the BF or FF surface as a function of the laser power density. The laser power density thresholds of the two cases are identified with vertical lines. (c and d) Representative optical images of nucleated surface bubble on (c) the BF surface and (d) the FF surface.

Our measurements show that the nucleation time can be shortened as the laser power density increases (**Figure 1b**). At the same time, however, it is found that at a certain power density, the nucleation times of bubbles on the FF surface are always shorter than those on the BF surface. We have also noticed that the thresholds of the power density to form a surface bubble in each of the two cases are drastically different. For nucleation on the FF surface, the threshold is $0.35\sim0.38\times10^6$ W/cm$^2$, but for the BF surface, the nucleation threshold is found to be much higher, $1.00\sim1.04\times10^6$ W/cm$^2$.

One mechanism that can potentially contribute to the surface bubble formation in the NP suspension is the volumetric photothermal heating(*28*), where the CS NPs in suspension absorb laser energy and heat up the laser-irradiated volume. However, if the bubble formation is such a purely thermal process, the nucleation dynamics would be similar for both the BF and FF surfaces when the laser is focused on them, respectively. Furthermore, if the surface bubble nucleation depends on the local temperature of the surface, as that in pool boiling, the threshold of laser power density for the bubble nucleation on the FF surface would be higher than on the BF surface, since the laser energy is attenuated by the light-absorbing NPs in suspension while reaching the FF surface. Using a finite element-based thermofluidic model (see Supporting Information SI2 for details), we have calculated the temperature distributions in the suspension when laser is focused on either surface, and found that the BF surface temperature is slightly higher (over 10 K) than the FF surface temperature under the same laser power density (Supporting Information, Figure S4). However, the observed bubble formations on the FF and BF surfaces show entirely opposite nucleation characteristics with bubble formation actually having a lower laser power density threshold on the FF surface (**Figure 1b**). Under the same laser power density, nucleation time of the bubble on the FF surface is also shorter than that on the BF surface (**Figures 1b, c** and **d**). Thus, the photothermal volumetric heating cannot explain the observed discrepancy in nucleation times on the BF and FF surfaces.

Upon detailed analysis of the side view high-speed videography, we observe that there are always CS NPs moving toward the surfaces leading up to every bubble nucleation (**Figures 2a and b**, Supporting Movies M3 and M4). In the experiments, we track the position of the glowing dots as a function of time, where the glowing dots correspond to the diffraction-limited scattered light from CS NPs. When the laser power density is low (< $0.35\times10^6$ W/cm$^2$), such NP motion along the laser beam axis is not apparent, and there is no bubble formation. When the laser is focused on the FF surface, NPs moving towards the surface is observed when the laser power density is above 0.35~$0.38\times10^6$ W/cm$^2$, and bubble nucleation follows. When the laser is focused on the BF surface, there are NPs moving towards the surface only if the laser power density is above 1.00~$1.04\times10^6$ W/cm$^2$, following which bubble nucleation is also observed. Such NP movements along the laser beam propagation direction are observed within ~50 μm from the surfaces, and they cannot be driven by thermal convective flow, since it is vertical near the walls of the upright cuvette (see Supporting Information, Figure S3). It thus evidents that NP moving toward the focused surface is a necessity for surface bubble nucleation. Then the question is what drives such NP movements?

Our previous studies have found that CS NPs in a suspension can be driven by dispersive optical scattering force originated from the momentum exchange between incident photons and the NPs (*42*, *43*). The photon stream in the laser beam usually exerts an optical pushing force that drives the CS NPs to move in the light propagating direction. However, as reported in several previous works(*43–50*), plasmonic vapor nanobubbles can be formed around the heated CS NPs irradiated by a pulsed laser at the SPR peak of the NPs. This supercavitation (i.e., nanobubble encapsulating the NP) can optically couple to the encapsulated NP to trigger the "negative" optical scattering forces on the NP, leading to an optical pulling force (**Figure 2c**), depending on the position of a CS NP inside the nanobubble (*42, 43*). It means that the laser beam can drive the CS NP to move against the photo stream, and this is why some NPs are seen moving against the light propagation direction towards the BF surface. Since such supercavitation needs relatively high laser power density to intensely heat the NP, the pulling motion is not observed until a laser power density threshold is reached. It is worth noting that the fluence to create the supercavitating nanobubble has been known to be ~7 mJ/cm$^2$ (*48*), which is close to the threshold of the laser power density (1.00~$1.04\times10^6$ W/cm$^2$, converted to fluence is 7.4~7.7 mJ/cm$^2$) to form the bubble on the BF surface. On the contrary, the CS NPs driven by the optical pushing force can occur without the need of supercavitation. Therefore, we see plenty of NPs moving toward the FF surface even with a laser power density of $0.37\times10^6$ W/cm$^2$. These facts lead us to believe that it is such optical forces that drive the NPs to be deposited on the surfaces, which then serve as the heating source on the surface for bubble nucleation.

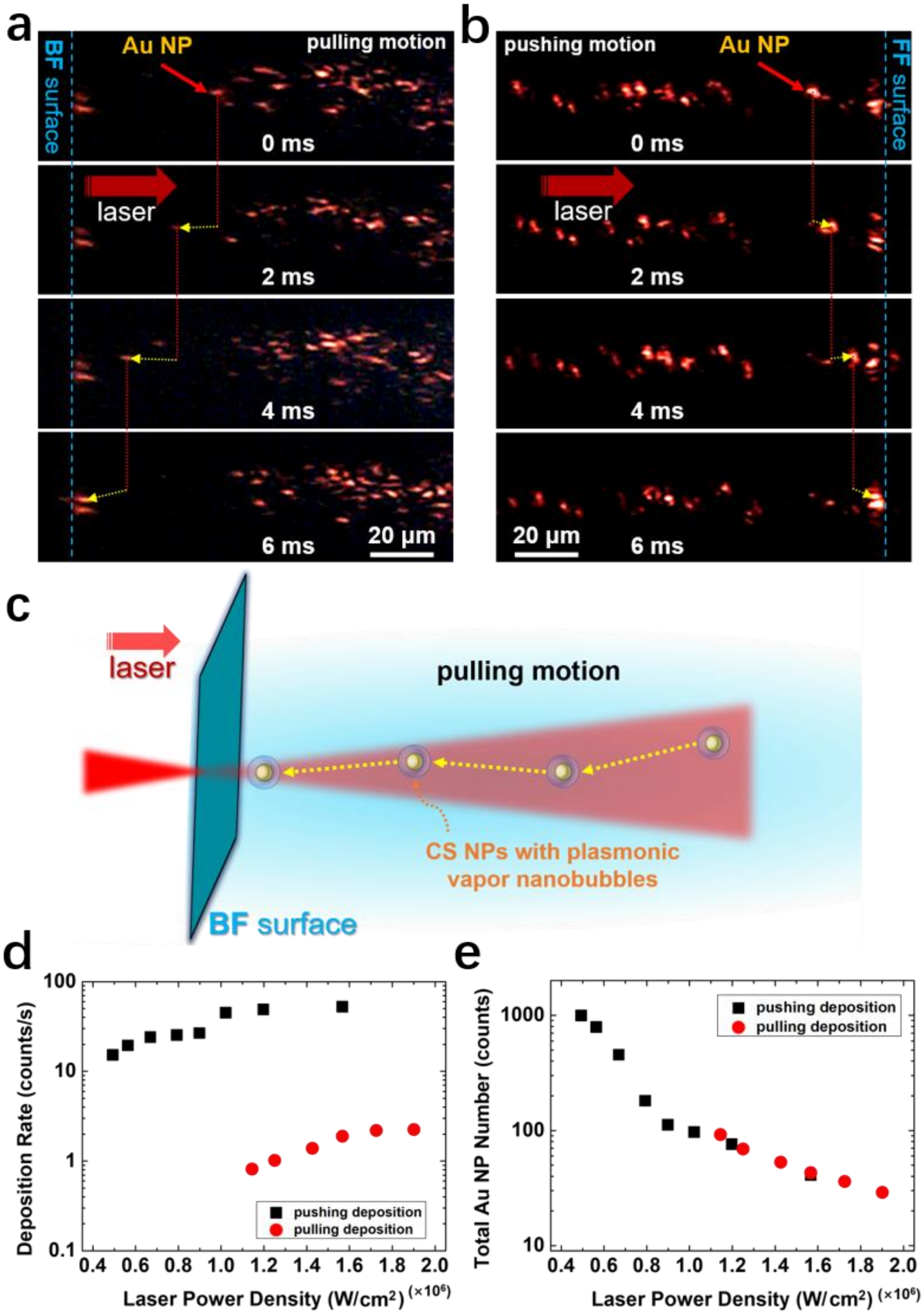

**Figure 2.** (a and b) Dark field optical images of optically driven CS NPs moving (a) against the light propagation direction by optical pulling force, and (b) along the light propagation direction by optical pushing force, as a function of time. The moving NPs are indicated by red arrows, and the yellow arrows show the trajectory of the NP between two frames. (c) Schematic of a supercavitating CS NP that has an optical condition enabling optical pulling motion (*42*, *43*). (d) The deposition rate of CS NP, and (e) the average total number of CS NP deposited on surface prior to surface bubble nucleation as a function of the laser power density.

To quantitatively investigate the relationship between the optically driven CS NPs and the surface bubble formation, we analyze the number of CS NPs moving towards and reaching either surface prior to bubble nucleation with the high-speed videography. It is found that the number of CS NP deposited on the BF surface per unit time (i.e., deposition rate) is one order of magnitude lower than that on the FF surface (see **Figure 2d**). This observation is reasonable as enabling the optical pulling of a CS NP requires the presence of a encapsulating nanobubble, and even with the supercavitation, the NP can only receive optical pulling force when it is in a certain region inside the nanobubble as previously studied in Ref. (*43*). This is in sharp contrast to the cases of optical pushing motion that happens without the need of supercavitation. Even with supercavitation, optical pushing of NP is more likely than pulling since there is a larger spatial portion inside the nanobubble where the NP will experience optical pushing force (*43*). As a result, there is a higher possibility that CS NPs undergo pushing motion than pulling motion under laser irradiation, which leads to the higher NP deposition rates on the FF surface. Interestingly, the accumulated numbers of CS NPs deposited on the BF and FF surfaces prior to bubble nucleation are almost the same for a certain power density range (**Figure 2e**). This strongly indicates that the bubble nucleation is due to the surface heating provided by the deposited NPs via the photothermal energy conversion. It also suggests that the surface temperature history is not important to nucleation, but it is the instantaneous photothermal heat flux when sufficient NPs are deposited on surface that triggers the bubble nucleation. That is to say, temperature profile around a deposited NP can reach steady state in a time scale much shorter than both the surface bubble nucleation and NP deposition time, and a threshold photothermal heat flux provided by the deposited NP needs to be met for bubble nucleation. When the laser power density increases, less deposited NPs should be needed to achieve bubble nucleation. This is evident in **Figure 2e**, which shows that the numbers of accumulated NPs both on the BF and FF surfaces prior to bubble nucleation decrease as the power density of the laser increases.

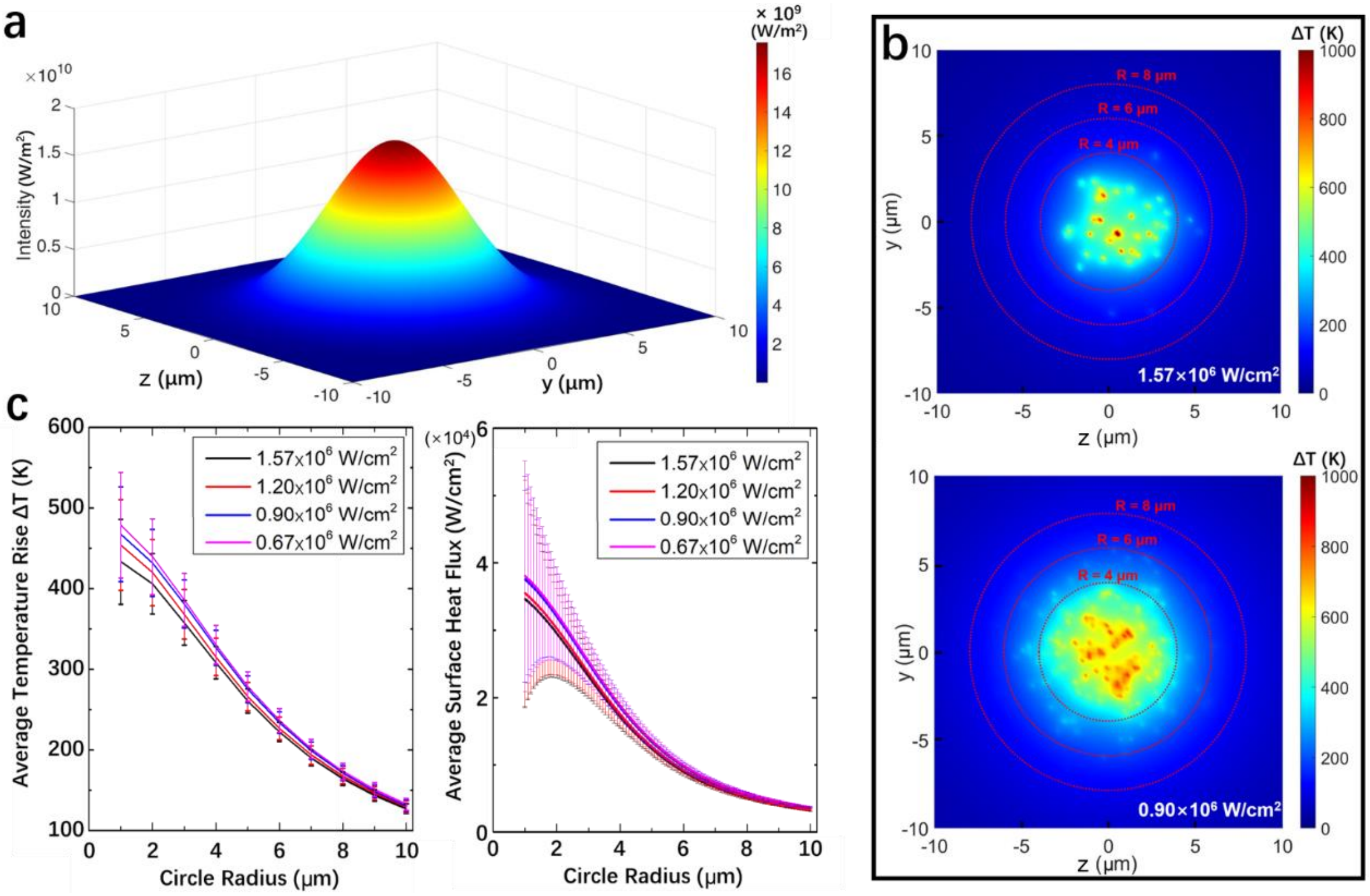

**Figure 3.** (a) Simulated incident laser power density profile, with the laser power density of 1.76×10$^6$ W/cm$^2$ (corresponds to the laser power of 1W). (b) Calculated surface temperature rise profiles in the NP deposition area under the laser power densities of (top panel) 1.57×10$^6$ W/cm$^2$ (corresponds to the laser power of 890 mW), and (bottom panel) 0.90×10$^6$ W/cm$^2$ (corresponds to the laser power of 510 mW). (c) Left panel: the average surface temperature rise (with respect to room temperature), and right panel: the average surface heat flux of the NP deposition area with different radii from laser beam center under different laser power densities. The circular area radii equal to 4, 6 and 8 μm are depicted in (b).

To investigate the thermal threshold to enable bubble nucleation, we study the average surface temperature rise from room temperature profile and heat flux of the area enclosing the deposited CS NPs under laser irradiation using Monte-Carlo (MC) simulations with the following assumptions. The location of a deposited NP on the surface is given by a probability function proportional to the intensity profile of the focused laser spot (i.e., Gaussian profile, see **Figure 3a**). The quality factor of the optical absorption of each deposited NP is equal to that of a single isolated NP on the substrate. The heat flux generated by a deposited NP equals to the absorbed optical power by the NP, and the light intensity irradiating on this NP is given by the intensity profile of the laser spot at the location of the NP. The temperature profile on the surface around a deposited NP reaches steady state in a time scale (10$^{-6}$~10$^{-8}$ seconds) (*30*) that is much shorter than the inverse of the deposition rate of the optically driven NPs (0.1~1 s, see **Figure 2d**). The details of the simulation can be found in the Supporting Information, SI3. Taking the experimentally measured number of deposited CS NPs prior to surface bubble nucleation (**Figure 2e**) as an input, the MC simulation is used to produce one hundred independent configurations of randomly deposited NPs for each laser power density. For each configuration, the resultant surface temperature profiles and heat flux (10,000 configurations for heat flux simulations) for this given laser power density is calculated (see Supporting Information, SI3). At the power density between $0.67\times10^6$ W/cm$^2$~$1.57\times10^6$ W/cm$^2$, it is found that there are local hot spots with temperatures rise (ΔT) up to ~1000 K, where multiple NPs are closely deposited (see **Figure 3b**). However, we believe the observed bubble nucleation is not driven by these hot spots, which strongly depends on the configuration of deposited NPs, since the numbers of deposited NPs in each of the five runs of the same experiment do not differ more than 10%. On the other hand, when we pick five random configurations from the MC simulation, the chance to obtain observe such hot spots is very small. In another word, the emergence of the hot spot strongly depends on the spatial distribution of the deposited NPs. As a result, we believe the bubble nucleation we observed is more related to the average thermal condition of the surface.

We then investigate the average surface temperature and heat flux in circular areas with different radii, as shown in **Figure 3c**. The areas are defined as the circles centered at the origin of the Gaussian beam profile (**Figure 3b**). The temperature profile and heat flux are integrated over the circular region and then divided by its area. In **Figure 3c**, the area-normalized average temperatures and heat flux at different laser power densities are plotted as a function of the circle radius. Both the calculated average temperature profiles and heat flux are within the error bars of each other for different laser power densities, and as the circle radius increases, the average temperature and heat flux from different laser power densities further converge. The average surface temperatures at the radius of ~10 μm are almost the same (~420 K, which is ΔT=130 K) regardless of the laser power density. Interestingly, this temperature is very similar to the critical nucleation temperature (422 K) of surface bubble on pre-deposited plasmonic surfaces reported in ref. [(*30*)]. The radius of ~10 μm is also close to the size of the initial giant bubble, based on which nucleation was defined in ref. [(*30*)]. Near the center of the circle, the average temperature is much higher than the critical nucleation temperature. This at least supports that the experimentally determined number of deposited NPs prior to bubble nucleation can lead to the surface temperatures high enough to form a surface bubble. These findings indicate that it is the surface heating effect from the deposited NP that leads to the surface bubble nucleation in a NP suspension. For a given laser power density, it is necessary to accumulate sufficient NPs on the surface to reach the nucleation temperature before a surface bubble can be formed. When the laser power density is lower, more NPs need to be deposited to convert the laser energy into heat to raise the temperature to the critical value for bubble nucleation, and when the laser power density is higher, less NPs need to be deposited to reach the same effect.

As a comparison, we also studied surface bubble nucleation in a suspension of solid Au NP. Each of the spherical solid Au NP has a diameter of 103 ± 10 nm. The concentration of the solid Au NPs is $\sim4.3\times10^{15}$ particles/m$^3$, and dissolved air level is the same as the CS Au NPs suspension (i.e., air equilibrium). We firstly focus the laser on the BF surface, but we cannot observe any pulling motion of the solid Au NPs (Supporting Movies M5). Because of the mismatch between the SPR wavelength of the solid Au NP (563 nm) and the incident laser wavelength (800 nm), there is no significant plasmonic heating effect, and thus it is very difficult for the NP to form supercavitation, which is essential for achieving the optical pulling effect (*42, 43*). In this case, no surface bubble can be formed on the BF surface even the laser power density is increased to $1.85\times10^6$ W/cm$^2$ (corresponds to the laser power of 1.05 W), which is the maximum that can be achieved in our experimental setup. However, we can still see robust optical pushing motion of NPs when the laser focal plane is on the FF surface (**Figure 4a**, Supporting Movie M6). This pushing motion again results in the deposition of NPs on the surface, and thus surface bubbles can still be formed and grow under the irradiation of the off-SPR laser beam (Supporting Movie M6). The bubble nucleation time as a function of laser power density is plotted in **Figure 4b**. This experiment further confirms that the NP deposition is a necessity for photothermal surface bubble formation in plasmonic suspensions.

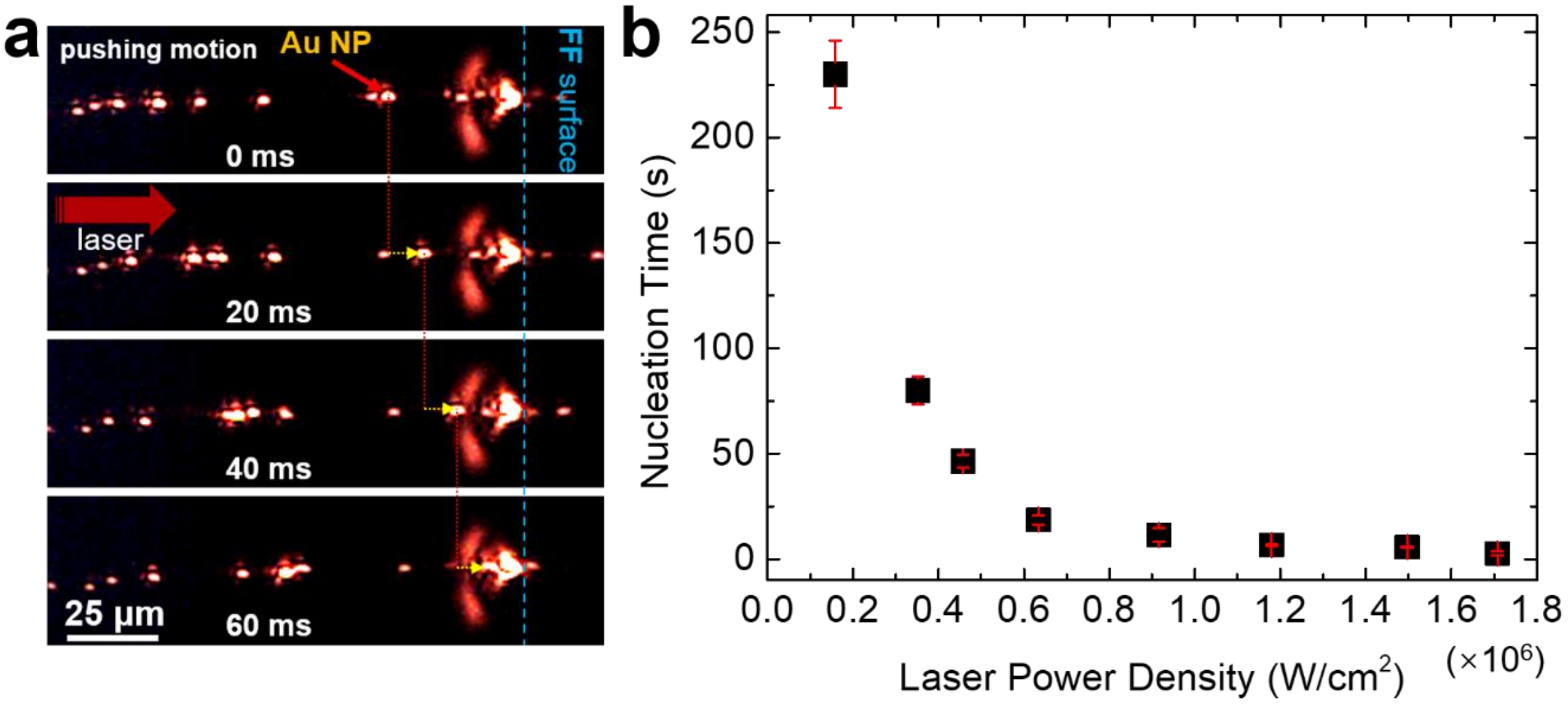

**Figure 4.** (a) Optical images of the optical pushing motion of a solid spherical Au NP indicated by the red arrow. The focal plane of the laser is on the FF surface. The time interval between each image is 20 ms. (b) Surface bubble nucleation time as a function of laser power density in the pushing case of solid spherical Au NP.

## CONCLUSIONS

In conclusion, we have elucidated the mechanism of surface bubble formation on a transparent bare quartz surface immersed in plasmonic CS or solid Au NP suspension. The forward- or backward-moving NPs driven by optical pushing or pulling force can be deposited on the surface and then act as surface photothermal plasmonic heaters on the transparent substrate. There is a critical number of deposited NPs at a given power density of the laser so that the surface heating effect can allow the surface to reach a threshold temperature for the nucleation of surface bubbles. Furthermore, bubble nucleation on the BF surface is only possible if the incident laser frequency coincides with the SPR peak of the NP since intense plasmonic heating is needed to generate a supercavitation – a necessity for optical pulling deposition.

## METHODS

**Optical system to generate and characterize the plasmonic surface bubble: Figure 1a** schematically shows the experimental setup used to analyze the nucleation of plasmonic surface bubbles. A femtosecond mode-locked monochromatic pulsed laser (repetition rate of 80.7 MHz and pulse duration of 200 fs) from a Ti:sapphire crystal in an optical cavity (Spectra Physics, Tsunami) is directed into a UV quartz cuvette (Alpha Nanotech Inc.) containing Au NP suspensions. The center wavelength of the laser is 800 nm with a full-width-half-maximum of ~10.5 nm. The beam is focused by a 20× objective lens (Edmund Optics) onto the inner wall of the cuvette with a $1/e^2$ radius of 6 μm. The laser power can be tuned continuously from 10 mW ($1.76\times10^4$ W/cm$^2$) to 1.05 W ($1.85\times10^6$ W/cm$^2$) using a continuously variable metallic neutral density filter (NDC-25C-4M, Thorlabs). An optical shutter controls the on/off of the laser. The four-side-polished UV quartz cuvette has a chamber with a width of 1 mm. Before containing Au NP suspensions (AuroShell, Nanospectra Biosciences, Inc.), the cuvette was sequentially cleaned with deionized (DI) water, acetone, isopropyl alcohol and ethanol in an ultrasonic bath. With respect to the laser incident direction, the two inner surfaces of the cuvette are labeled as backward-facing (BF) and forward-facing (FF) surfaces (**Figure 1a**). Core-shell (CS) Au NPs consisting of a ~50 nm silica core and a ~10 nm Au shell are dispersed in DI water at a concentration of ~$2\times10^{15}$ particles/m$^3$. The SPR peak of this CS Au NPs in the suspension is around 780~800 nm, which coincides with the laser wavelength. In the comparison experiments, the spherical solid Au NP (the SPR wavelength is 563 nm) used has a diameter of 103 ± 10 nm, and the concentration is ~$4.3\times10^{15}$ particles/m$^3$. A high-speed digital camera (HX-7, NAC) with a 10× objective lens (Edmund Optics) and a white LED (300 lm) illumination source are employed to record the nucleation of surface bubbles from the side view.

## ACKNOWLEDGEMENTS

This work is supported by National Science Foundation (1706039) and the Center for the Advancement of Science in Space (GA-2018-268). T.L. would also like to thank the support from the Dorini Family endowed professorship in energy studies.

## AUTHOR CONTRIBUTIONS

Q.Z., E.L. and T.L. designed the experiments, and Q.Z. set up the experiments. Q.Z. performed the experiment. Q.Z., R.L., E.L. and T.L. designed the simulations and Q.Z. performed the simulations. Q.Z. wrote the manuscript, R.L., E.L. and T.L. revised it.

## COMPETING INTERESTS

The authors declare no conflict of interest.

# Supporting Information for

# Optically Driven Gold Nanoparticles Seed Surface Bubble Nucleation in Plasmonic Suspension

Qiushi Zhang[1], Ruiyang Li[1], Eungkyu Lee[1,4*], and Tengfei Luo[1,2,3*]

[1]Department of Aerospace and Mechanical Engineering, University of Notre Dame, Notre Dame, IN, USA

[2]Department of Chemical and Biomolecular Engineering, University of Notre Dame, Notre Dame, USA

[3]Center for Sustainable Energy of Notre Dame (ND Energy), University of Notre Dame, Notre Dame, USA

[4]Department of Mechanical System Engineering, Kumoh National Institute of Technology, 61 Daehak-ro, Gumi, Gyeongbuk, 730-701, South Korea

*Correspondence to: tluo@nd.edu; elee@kumoh.ac.kr

## SI1. Characterizing the CS NP deposition and surface bubble nucleation processes

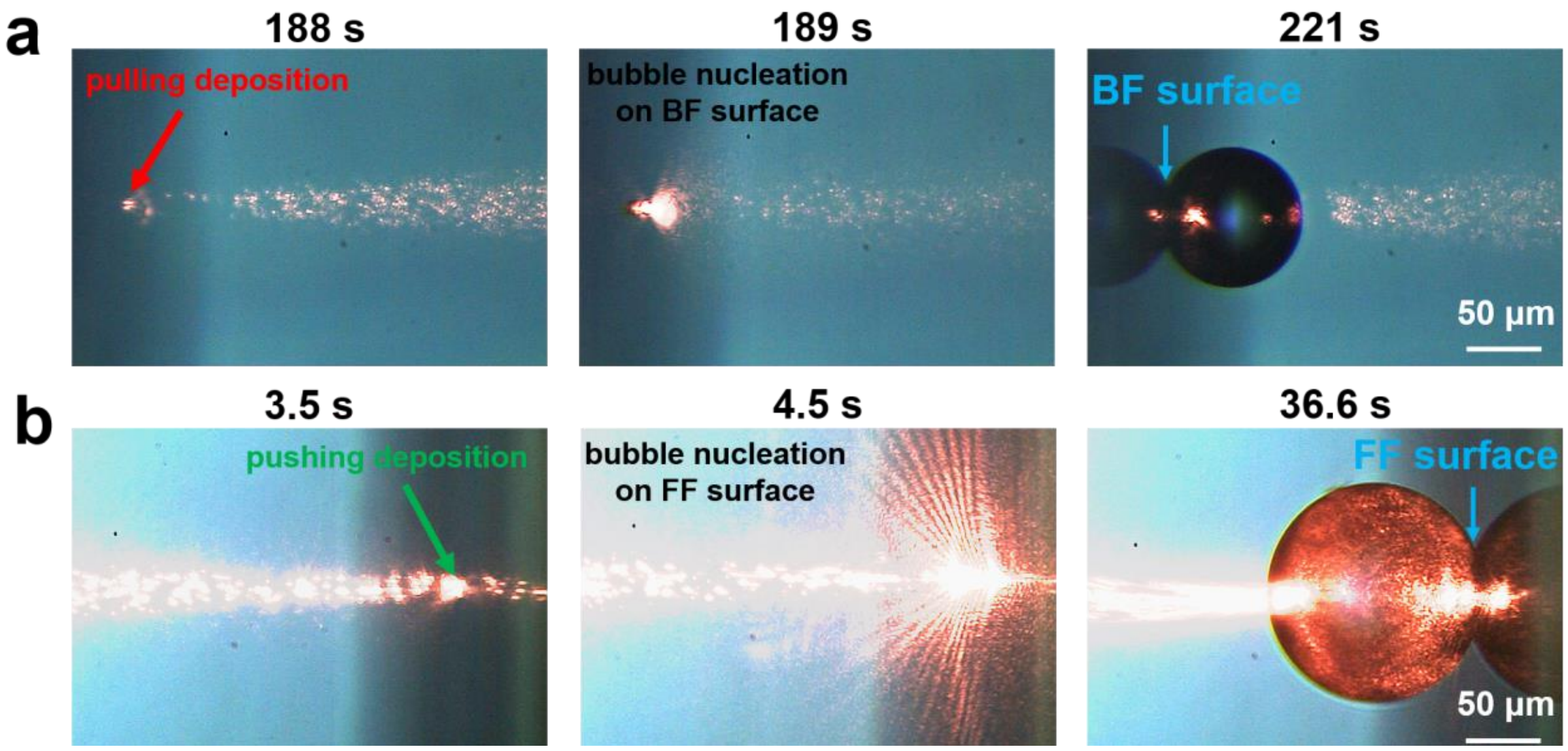

**Figure S1.** The optical images showing the CS NPs (the glowing dots indicated by a red/green arrow on the quartz surface) deposited by pulling/pushing deposition (left panels), the moment of surface bubble nucleation (middle panels) and the nucleated surface bubble (right panels) with the focal plane of laser on BF (a) or FF (b) surface. The laser powers for (a) and (b) are 0.58 W and 0.47 W, respectively. The time of each panel is indicated in the figure. The starting time is set to the moment of turning on the incident laser in (a) and (b).

## SI2. Finite element thermofluidic simulations details

We employed COMSOL Multiphysics to simulate the temperature and flow profiles of the CS NP suspension under the irradiation of incident laser. The flow effect, thermal conduction, the attenuation effect of NP suspension and thermal convection in liquid are included in our simulations. The geometry and mesh structures used in our simulations are shown in Figure S2. As can be seen, the geometry includes a layer of water sandwiched by two $SiO_2$ walls, which serve as the cuvette surfaces. The dimension of each component is similar to the experimental system. The incident laser is propagating along the x-direction and gravity is in the y-direction. According to the laser propagation direction, the BF and FF surfaces of quartz cuvette are identified, respectively. The insert zooms in the cylindrical region heated by laser, where we employed much finer mesh structures in simulation. As discussed in refs. [28, 29]*, the incident laser induces volumetric heating of the irradiation region, and the heat generation rate decays along the laser propagation direction by the attenuation factor of the Au NP suspension simulated.

* References are the same as those in the main text.

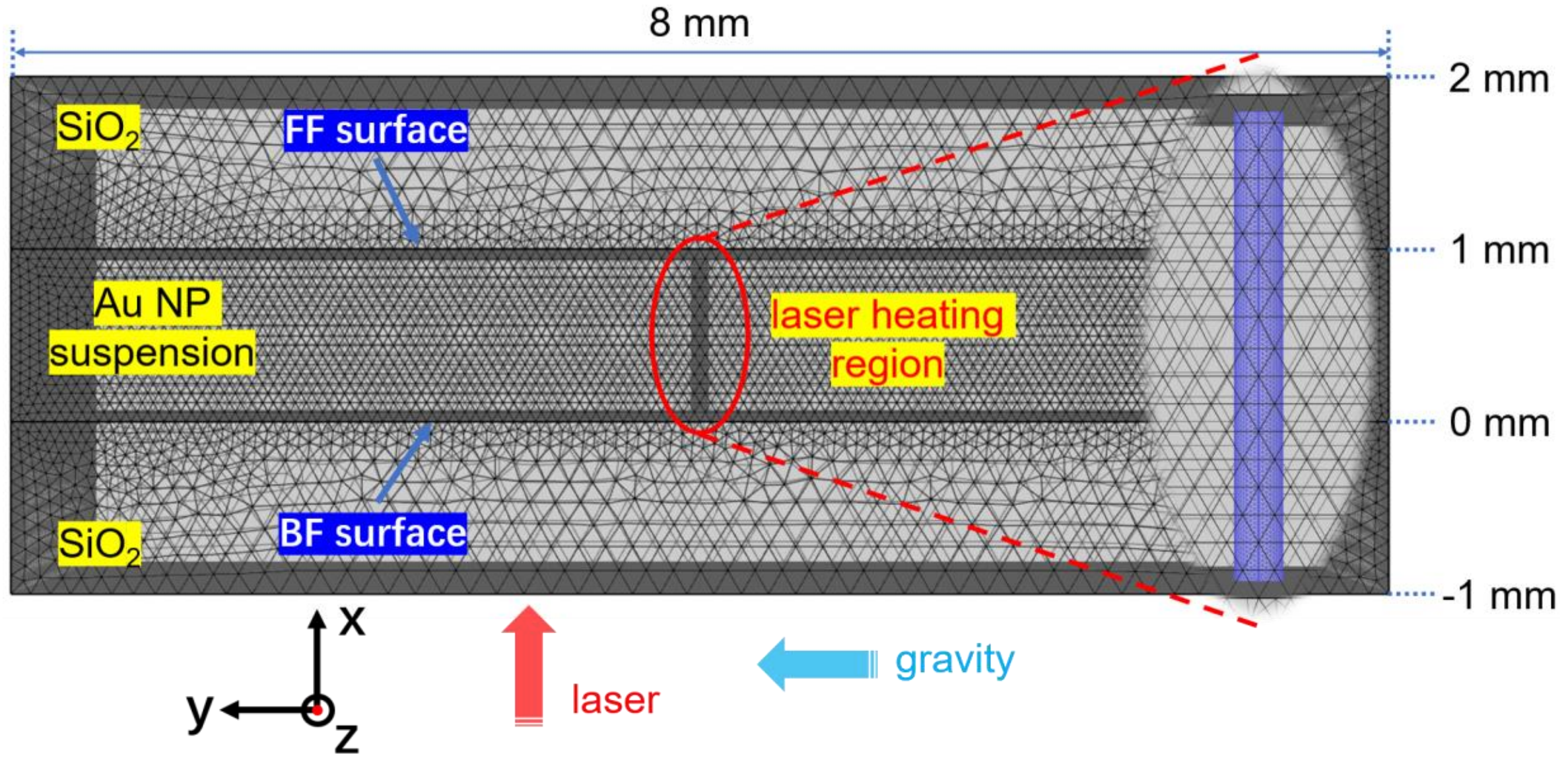

**Figure S2.** The geometry and mesh structure used in the thermofluidic simulations of the temperature and flow profiles in CS NP suspension upon laser irradiation. The insert zooms in the cylindrical region heated by the laser.

There are several conditions that have been assumed in our simulations: (1) The liquid flow and heat transfer are both transient (see Figures S3a and b). (2) In the liquid water, the flow is laminar and compressible with gravity perpendicular to the laser propagation direction (see Figure S2), which satisfies the following momentum equation:

$$\rho\frac{\partial\vec{u}}{\partial t}+\ \rho(\vec{u}\cdot\nabla)\vec{u}-\nabla\cdot\left(\mu(\overleftrightarrow{\nabla u}+\overleftrightarrow{\nabla u}^{\mathrm{T}})-\frac{2}{3}\mu(\nabla\cdot\vec{u})\overleftrightarrow{I}-p\overleftrightarrow{I}\right)-\ \rho\vec{g}=0 \tag{s1}$$

and continuity equation:

$$\frac{\partial\rho}{\partial t}+\ \nabla\cdot(\rho\vec{u})=0 \tag{s2}$$

where $\rho$ is the density of water, $\mu$ is the dynamic viscosity of water, $\vec{u}$ is the velocity vector, $p$ is pressure, $t$ is time, $\vec{g}$ is gravity constant, and $\overleftrightarrow{I}$ is a 3×3 identity matrix. (3) The quartz substrates are considered as rigid solid materials. (4) The volumetric heating [28, 29] ($Q_v$) of the Au NPs irradiated by laser is the only heat source, which supplies the heat to the liquid water with the following heat transfer equations:
In water,

$$\rho C_p\frac{\partial T}{\partial t}+\rho C_p\vec{u}\cdot\nabla T-k_w\nabla^2T=Q_v \tag{s3}$$

where $C_p$ is the heat capacity of water at constant pressure, $T$ is the temperature, $k_w$ is the thermal conductivity of water, $Q_v$ is the heat generation rate by the volumetric heating, and in the medium of quartz,

$$-k_s \nabla T = q \tag{s4}$$

where $k_s$ is the thermal conductivity of quartz and $q$ is the heat flux coming through the water/quartz boundary. The boundary conditions used in our simulations are similar to those in ref. [28]. The heat generation rate of the volumetric heating is as the following [28, 29]:

$$pulling\ case: Q_v = \eta_{abs} \cdot \alpha \cdot e^{-\alpha x} \cdot \frac{P_L}{2\pi\sigma^2[1+(\frac{\lambda x}{4\pi n\sigma^2})^2]} \cdot exp\left[-\left(\frac{y^2+z^2}{2\sigma^2[1+(\frac{\lambda x}{4\pi n\sigma^2})^2]}\right)\right] \tag{s5}$$

$$pushing\ case: Q_v = \eta_{abs} \cdot \alpha \cdot e^{-\alpha x} \cdot \frac{P_L}{2\pi\sigma^2[1+(\frac{\lambda(x-h)}{4\pi n\sigma^2})^2]} \cdot exp\left[-\left(\frac{y^2+z^2}{2\sigma^2[1+(\frac{\lambda(x-h)}{4\pi n\sigma^2})^2]}\right)\right] \tag{s6}$$

where $P_L$ is the incident laser power, and $\eta_{abs} \sim 0.2$ is the optical absorption efficiency of CS NPs, which is determined by the ratio of the absorption quality factor and the extinction quality factor of the CS NP in DI water [28]. $\lambda$ is the wavelength of the laser (~800 nm) and $n$ is the refraction index of water (~1.33). The optical attenuation factor of the NPs suspension $\alpha$ is ~ 262 $m^{-1}$, which is extracted from the absorbance spectrum [28]. For our 20× objective lens, $\sigma$ = 3 μm is the standard derivation of the Gaussian beam, and $h$ is the position of the laser focal plane in the x-direction, which is 0 in the pulling case and 1 mm in the pushing case. As we can see in equations (s5) and (s6), the heat generation rate ($Q_v$) for a given location is determined by the radius of the Gaussian distribution of laser intensity along x direction and in y-z plane (see Figure S2 for coordination definitions). Additionally, $Q_v$ is also influenced by the attenuation effect of the CS NPs in suspension along the propagation direction (x-axis).

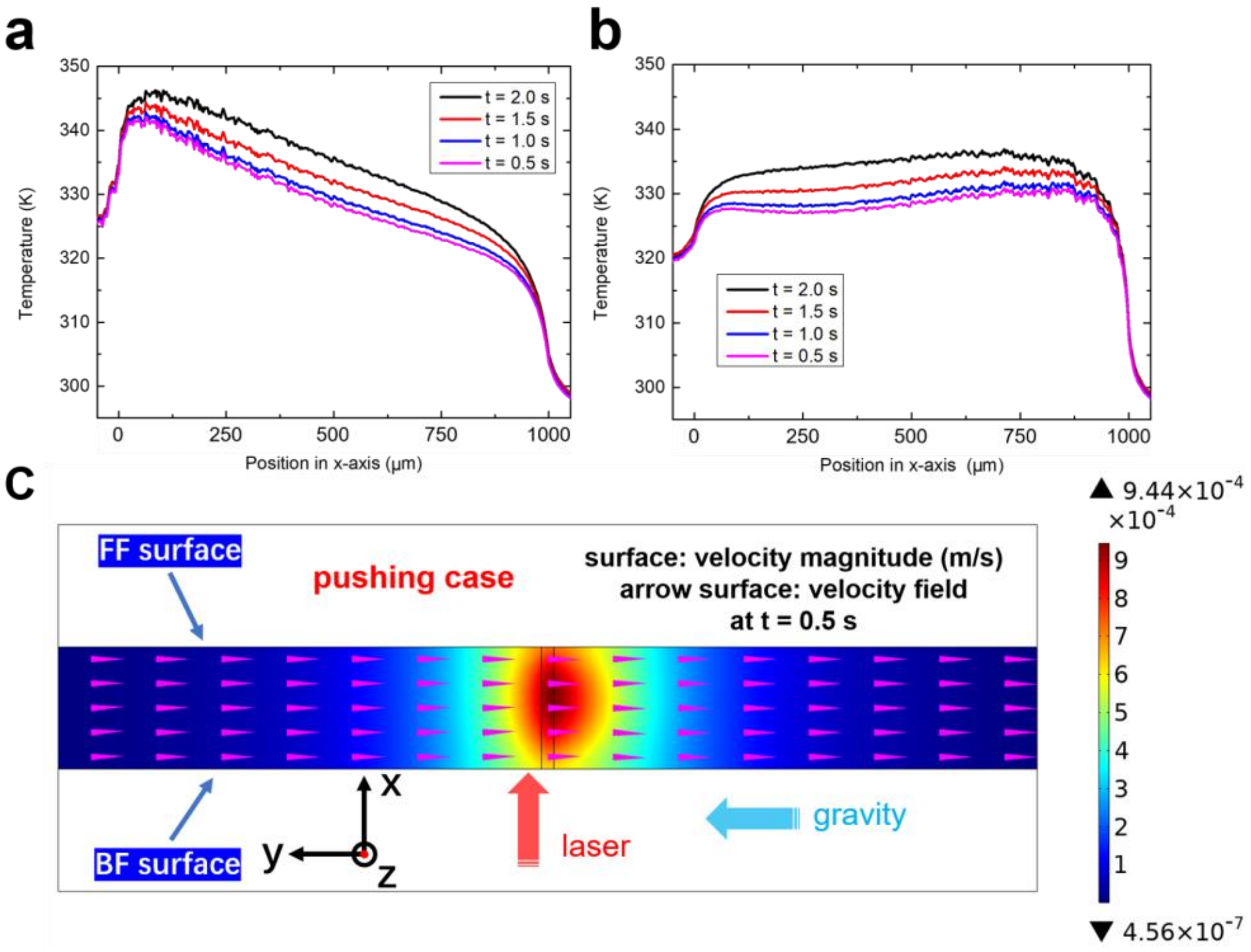

**Figure S3.** The calculated temperature profiles along the center axis of laser beam (in the x-direction) at different times in both pulling (a) and pushing (b) cases. (c) The simulated flow velocity map and vector field in the CS NP suspension of the pushing case (at $t$ = 0.5 s).

Figures S3a and b present the calculated temperature profiles along the center axis of the laser beam at different times in both pulling and pushing cases. As can be seen, the temperature of the NP suspension increases with time. However, the opposite symmetry between pulling and pushing cases always exists, and the overall temperature of the pulling case is higher than that of pushing case at any time, which is due to the laser attenuation effect of the NP suspension. Since the flow velocity maps of the pulling and pushing cases are similar, we only plot it for pushing case as an example. Figure S3c shows the flow velocity of the NP suspension in the pushing case. Due to the heating effect, the flow direction, as indicated by the magenta arrows, is opposite to gravity. The magnitude of flow velocity is much higher in the region irradiated by laser, which has higher local temperature. The simulated temperature distributions in the suspension when laser is focused on either surface are shown in Figure S4. It is clear that the local temperature is higher around the focal planes on BF and FF surfaces for the pulling and pushing cases, accordingly, and BF surface has higher temperature than FF surface (Figures 4c, d and e).

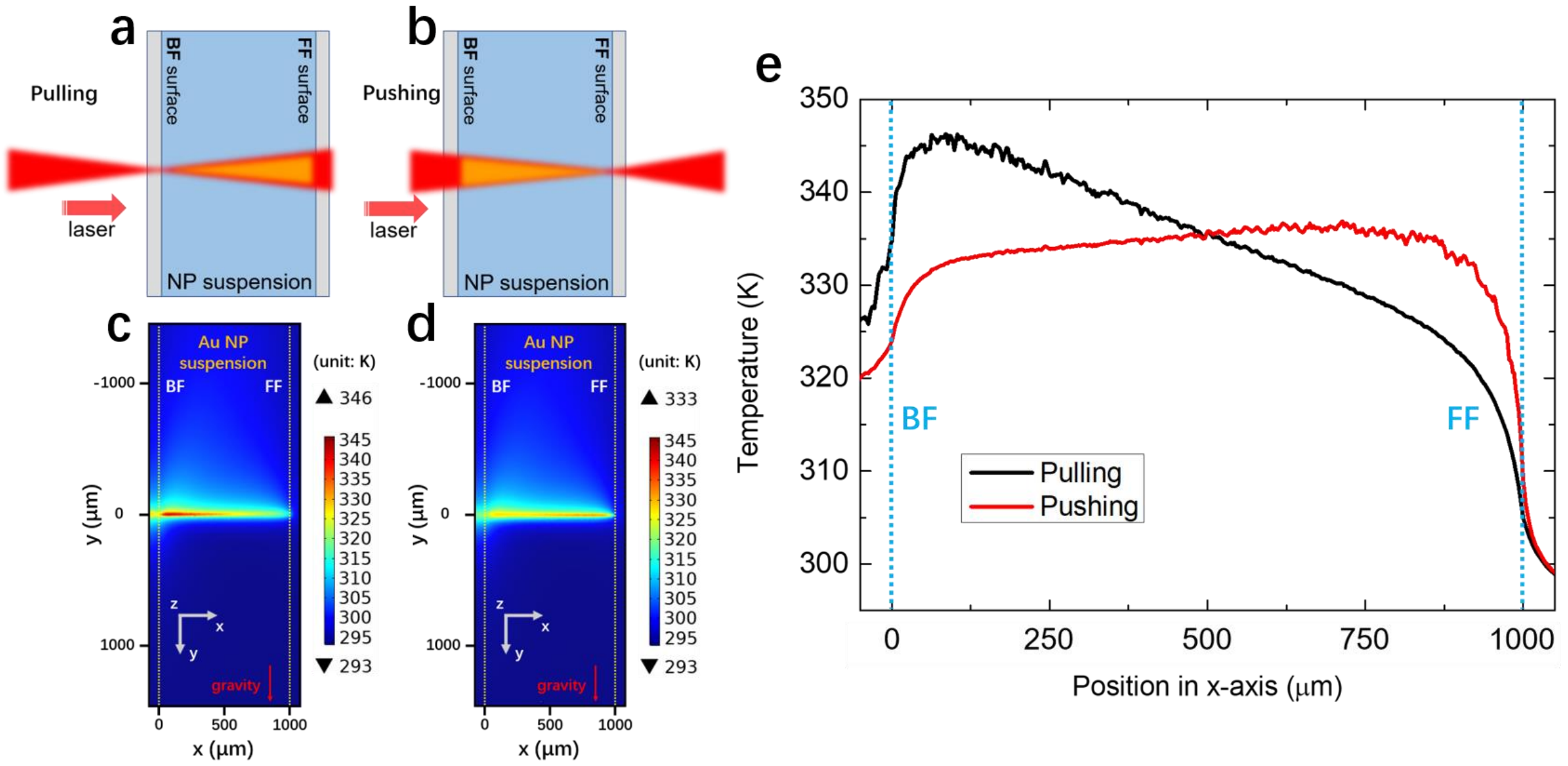

**Figure S4.** Schematic descriptions of the pulling and pushing cases with the focal plane of laser on the BF (a) and FF (b) surfaces. The temperature profiles on the plane crossing the center axis of laser beam from the thermofluidic simulations of the pulling (c) and pushing (d) cases, at t = 2.0 s. (e) The calculated temperature profiles along the center axis of the laser beam (in the x-direction) in both pulling (black) and pushing (red) cases, at t = 2.0 s.

## SI3. Monte-Carlo-assisted surface temperature field and heat flux simulations

To investigate the thermal threshold to enable bubble nucleation, we study the surface temperature profile and heat flux of the area enclosing the deposited CS NPs under laser irradiation using Monte-Carlo (MC) simulations with the following assumptions. The location of a deposited NP on the surface is given by a probability function proportional to the intensity profile of the focused laser spot (i.e., Gaussian profile, see **Figure 3a** in main text). The quality factor of the optical absorption of each deposited NP is equal to that of a single isolated NP on the substrate. The heat flux generated by a deposited NP equals to the absorbed optical power by the NP, and the light intensity irradiating on this NP is given by the intensity profile of the laser spot at the location of the NP. The temperature profile on the surface around a deposited NP reaches steady state in a time scale ($10^{-6}$~$10^{-8}$ seconds) [30] that is much shorter than the inverse of the deposition rate of the optically driven NPs (0.1~1 s, see **Figure 2d** in main text). With the total numbers of CS NP needed for surface bubble nucleation in the corresponding laser power densities (**Figure 2e** in main text), we then move to analyze the surface temperature profile and heat flux at the NP deposition area under the irradiation of the incident laser. The incident laser is a Gaussian beam with a $1/e^2$ radius of 6 μm. The laser power density (*I*) at a given location (*y*, *z*) can be described by:

$$I\ (y,z) = I_0 \times e^{-\frac{y^2+z^2}{2\sigma^2}} \qquad \text{(s7)}$$

where $I_0$ is the maximum laser power density which locates at the center of laser beam and $\sigma$ is the standard deviation of the Gaussian laser beam which equals to 3 μm. By integrating equation (s7) in the *y*-*z* space, we will obtain the incident laser power ($P_L$) as:

$$P_L = \iint_{-\infty}^{+\infty} I_0 \times e^{-\frac{y^2+z^2}{2\sigma^2}}\, dydz \qquad \text{(s8)}$$

Combining equations (s7) and (s8), we can calculate the maximum laser power density ($I_0$) and the expression of power density for a given laser power (the magnitude of laser power is measured by a laser power meter). A demonstration of incident laser power density profile with the laser power of 1 W is shown in **Figure 3a** in main text. Then, the electric field strength (*E*) of the incident laser beam can be calculated from the power density (*I*) by:

$$E = \sqrt{\frac{2I}{nc\varepsilon_0}} \qquad \text{(s9)}$$

where $n$ is the refractive index, $c$ is the speed of light in vacuum and $\varepsilon_0$ is the electrical permittivity. The electric field strength ($E_0$) also locates at the center of laser beam, where has the highest power density ($I_0$). By inputting the value of $E_0$ as the amplitude of incident laser electric field expression, we performed full-wave electromagnetic calculations [29] with finite element method to estimate the maximum dissipation power ($P_0$) of a single CS NP (see Figure S5 for schematic setup of the full-wave electromagnetic calculation).

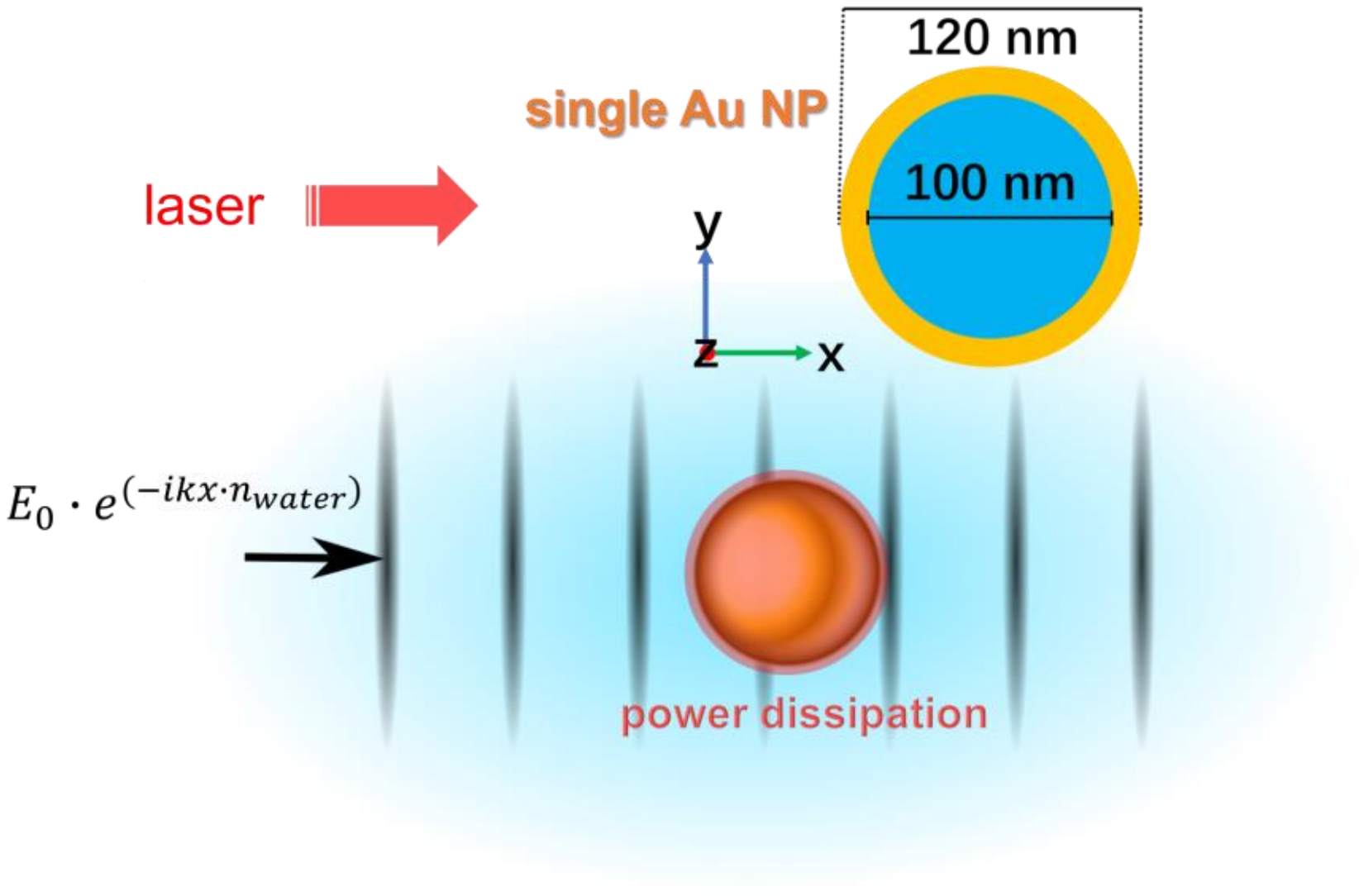

**Figure S5.** The schematic diagram of the full-wave electromagnetic finite element simulations. Laser propagates in the x-direction with the amplitude of electric field strength ($E_0$) calculated from (s9). The total dissipation power of a single Au NP is calculated from the volume integration. More details about this full-wave electromagnetic finite element simulation can be found in ref. [29].

Here, since we are using the maximum electric field strength ($E_0$) in this simulation, the simulated dissipation power corresponds to the maximum power that can be dissipated by a single Au NP under a given laser power. Next, we employ finite element heat conduction simulation [28] (convection is neglected at short time and length scales [30]) to calculate the temperature profile on the surface of quartz cuvette due to the heating of a single Au NP. Since heating up a single NP by plasmonic effect is much faster than the heat conduction process, we thus choose to simulate the steady state temperature profile [28, 29]. The geometry used in our simulation is shown in Figure S6a. A layer of water is sandwiched by two slices of $SiO_2$ substrates. A single NP is immerged inside the water layer. The dimensions of the system are large enough to eliminate the size effect (three orders of magnitude larger than the size of NP). The NP is the only heat source of the whole system, whose heat generation rate is calculated by:

$$Q_b = \frac{P_0}{V} \quad \text{(s10)}$$

where $P_0$ is the simulated maximum dissipation power and $V$ is the volume of a single NP. The NP sits on the surface of the bottom $SiO_2$ substrate with the NP/substrate contacting point shown in Figure S6a. Because of the heat transfer from the NP, the temperature of the surface of the bottom $SiO_2$ substrate will increase.

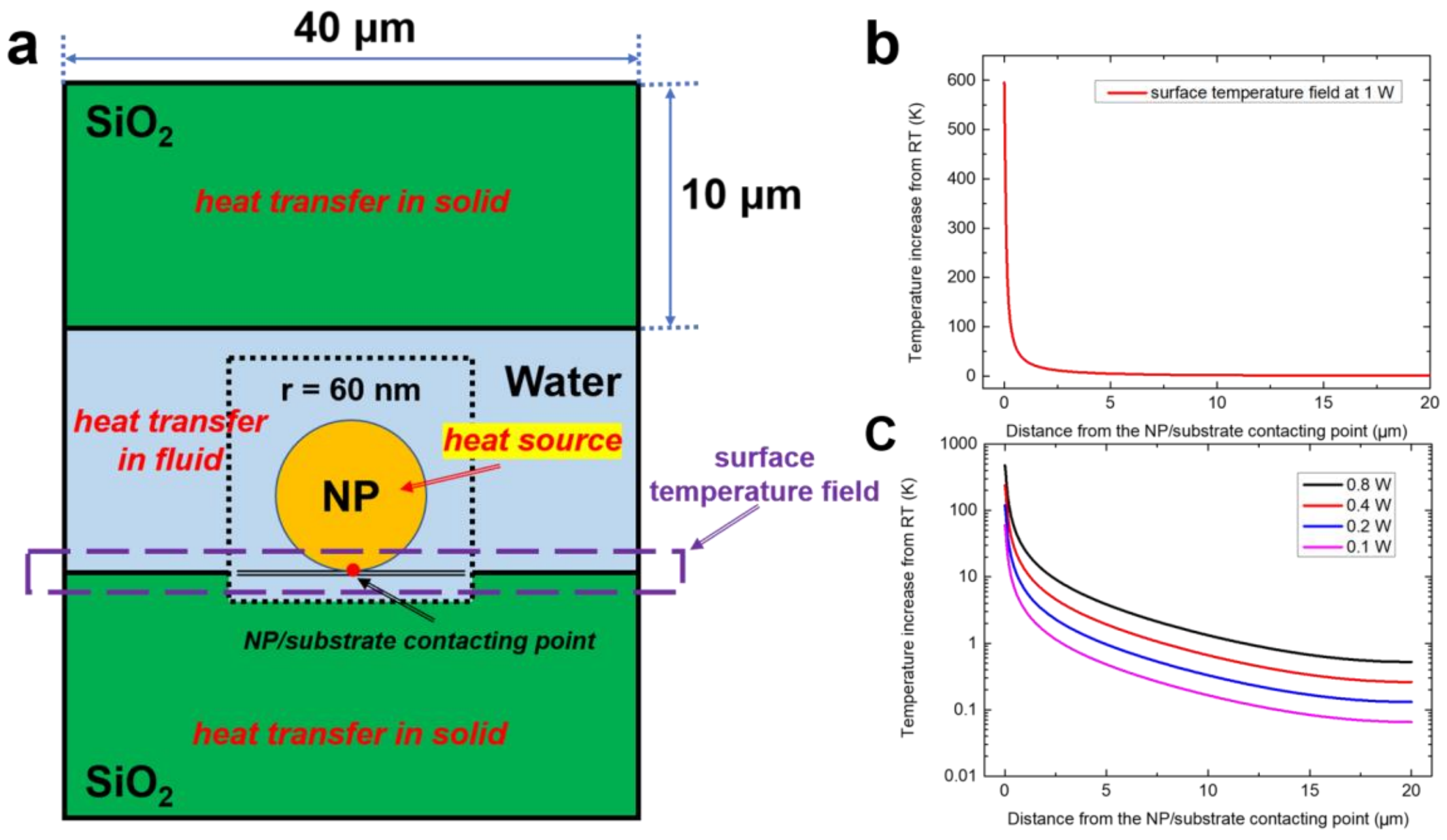

**Figure S6.** (a) The schematic diagram of the single Au NP-on-surface heat conduction simulations. (b) A sample simulated surface temperature profile with the incident laser power of 1 W. (c) Surface temperature profiles (in log scale) for laser powers of 0.1, 0.2, 0.4 and 0.8 W.

With all these simulations discussed above, we can finally calculate the temperature profile on the surface of quartz cuvette and heat flux generated by a single CS NP for a given laser power. A sample simulated temperature profile with the incident laser power equaling to 1 W is shown in Figure S6b. One thing to note is that this simulated temperature profile is the highest temperature can be generated by a single NP for a given laser power (since we use the maximum dissipation power), which means this single NP is assumed to locate at the center of laser beam. Figure S6c plots the maximum surface temperature profiles generated by the single NP under different incident laser powers. It is interesting to see that the maximum surface temperature field is linear to the magnitude of incident laser power. With this linearity, we can quickly obtain the temperature profile and heat flux of the surface deposited with multiple NPs. We then calculate the entire surface temperature profile and heat flux of the many CS NP-deposited area prior to bubble nucleation under a given laser power: Firstly, we use MC simulations to randomly deposit many CS NPs on site according to the probability function of laser power density which follows a Gaussian distribution ($\sigma$ = 3 μm). The total number of NP deposited in the simulation of a given laser power is determined from the experimental data, as shown in **Figure 2e** in main text, which is the total number of NP needed for each nucleation process. Secondly, the laser power density distribution and maximum surface temperature field generated by the single NP for an incident laser power can be obtained by the methods described above. Then, we can calculate the power density of each deposited NP according to the distance between this NP and the center of laser beam. Once we know the power density at the location of each NP, we can use the EM simulation as described above to calculate the dissipation power of the NP. The surface averaged heat flux (**Figure 3c** in main text, right panel) is obtained by adding up the dissipation power of each individual NP and dividing the total dissipation power by the surface area of each circle with a different radius. Because the surface temperature field generated by individual NP is linearly related to incident laser power and power density, we can obtain the surface temperature field of each deposited NP from the maximum surface temperature field scaled by the ratio of the power density at the location of the corresponding NP to the maximum power density for this laser power. With the surface temperature field of each deposited NP calculated, we will finally be able to simulate the temperature profile of the entire NP-deposited area by the superposition of individual NP-induced temperature fields. Two samples showing simulated surface temperature profiles of the entire NP-deposited area are presented in **Figure 3b** in main text.